\documentclass[journal=jacsat,manuscript=article]{achemso}

\usepackage[version=3]{mhchem} 
\usepackage{amssymb}

\author{Shunsuke Murai}
\affiliation{Department of Material Chemistry, Graduate School of Engineering, Kyoto University, Nishikyo-ku,
Kyoto 615-8510, Japan}
\alsoaffiliation{Department of Applied Physics and Institute for Photonic Integration, Eindhoven University of Technology, P.O. Box 513, 5600 MB Eindhoven, The Netherlands}
\email{murai@dipole7.kuic.kyoto-u.ac.jp}
\author{Diego R. Abujetas}
\affiliation{Physics Department, Fribourg University, Chemin de Musée 3, 1700 Fribourg Switzerland}
\author{Libei Liu}
\affiliation{Department of Material Chemistry, Graduate School of Engineering, Kyoto University, Nishikyo-ku,
Kyoto 615-8510, Japan}
\author{Gabriel W. Castellanos}
\affiliation{Department of Applied Physics and Institute for Photonic Integration, Eindhoven University of Technology, P.O. Box 513, 5600 MB Eindhoven, The Netherlands}
\author{Vincenzo Giannini}
\affiliation{Instituto de Estructura de la Materia (IEM-CSIC), Consejo Superior de Investigaciones Científicas, Serrano 121, 28006 Madrid, Spain}
\alsoaffiliation{Centre of Excellence ENSEMBLE3 sp. z o.o., Wolczynska 133, Warsaw, 01-919, Poland}
\alsoaffiliation{Technology Innovation Institute, Masdar City 9639, Abu Dhabi, United Arab Emirates}
\email{v.giannini@csic.es}
\author{Jos\'e A. S\'anchez-Gil}
\affiliation{Instituto de Estructura de la Materia (IEM-CSIC), Consejo Superior de Investigaciones Científicas, Serrano 121, 28006 Madrid, Spain}
\email{j.sanchez@csic.es}
\author{Katsuhisa Tanaka}
\affiliation{Department of Material Chemistry, Graduate School of Engineering, Kyoto University, Nishikyo-ku, Kyoto 615-8510, Japan}
\author{Jaime Gómez Rivas}
\affiliation{Department of Applied Physics and Institute for Photonic Integration, Eindhoven University of Technology, P.O. Box 513, 5600 MB Eindhoven, The Netherlands}
\email{j.gomez.rivas@tue.nl}

\title[An \textsf{achemso} demo]
  {Engineering bound states in the continuum at telecom wavelengths with non-Bravais lattices}

\abbreviations{IR,NMR,UV}

\begin{document}


\begin{tocentry}
\includegraphics{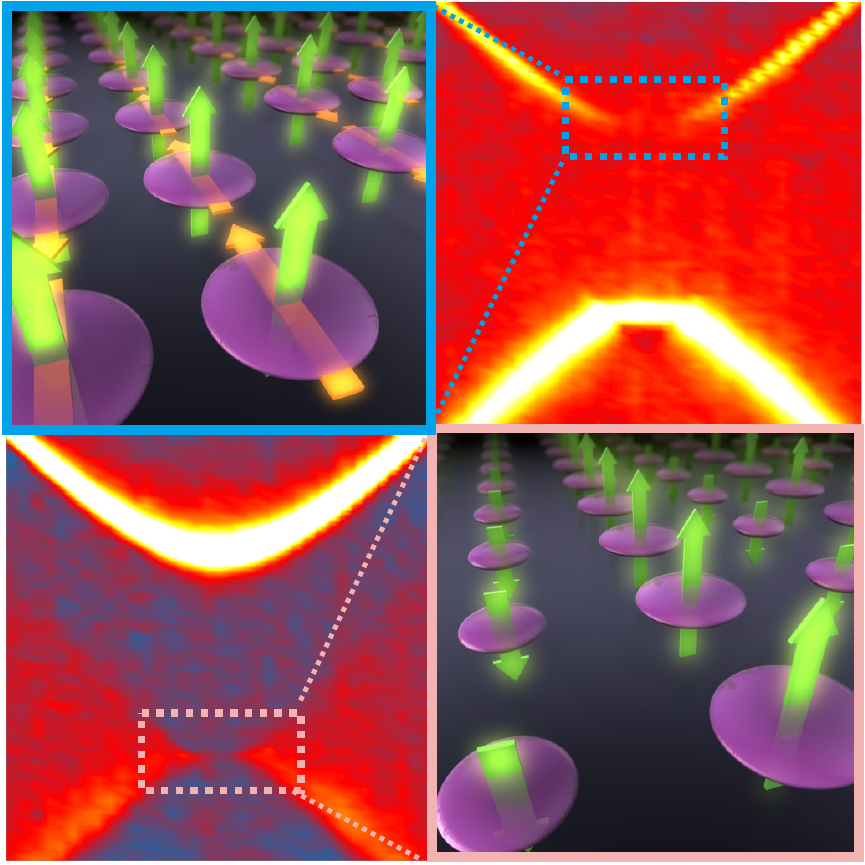}
\end{tocentry}




\begin{abstract}
 Various optical phenomena can be induced in periodic arrays of nanoparticles by the radiative coupling of the local dipoles in each particle. Probably the most impressive example is bound states in the continuum (BICs), which are electromagnetic modes with a dispersion inside the light cone but infinite lifetime, i.e., modes that cannot leak to the continuum. Symmetry-protected BICs appear at highly symmetric points in the dispersion of periodic systems. Although the addition of nonequivalent lattice points in a unit cell is an easy and straightforward way of tuning the symmetry, BICs in such particle lattice, i.e., non-Bravais lattice, are less explored among periodic systems. Starting from a periodic square lattice of Si nanodisks, we have prepared three non-Bravais lattices by detuning size and position of the second disk in the unit cell. Diffraction-induced coupling excites magnetic/electric dipoles in each nanodisk, producing two surface lattice resonances at the $\Gamma$ point with a band gap in between. 
 The high/low energy branch becomes a BIC for the size/position-detuned array, respectively, while both branches are bright (or leaky) when both size and position are detuned simultaneously. The role of magnetic and electric resonances in dielectric nanoparticles and the change of BIC to bright character of the modes is explained by the two different origins of BICs in the detuned arrays, which is further discussed with the aid of a coupled electric and magnetic dipole model. This study gives a simple way of tuning BICs at telecom wavelengths in non-Bravais lattices, including both plasmonic and dielectric systems, thus scalable to a wide range of frequencies.
\end{abstract}

\section{Introduction}
Periodic arrays of resonant particles represent an arena of intensive research in modern optics. Localized resonances at each particle can be radiatively coupled via in-plane diffraction, causing hybridized modes that are called surface lattice resonances (SLRs)\cite{RN216,RN648,RN443,RN145,RN138,RN224,RN483,RN484,RN264,RN2416,RN1291,RN141}. SLRs are characterized by narrow and strong resonances that appear in a wide spectrum range from terahertz\cite{RN3067} to ultraviolet\cite{RN773}, depending on the periodicity of the lattice and the size of the particles. 
The arrays include one dimensional chains\cite{RN443,RN648,RN3060} and two dimensional lattices of particles. The most-studied two dimensional lattices are square and hexagonal patterns, which fall into two of the five Bravais lattices in two dimensions. In a Bravais lattice, all the particles are crystallographically equivalent and can be generated by a set of translation operations. Some works have explored non-Bravais lattices, where a second set of basis is introduced to the initial Bravais lattice\cite{RN3038,RN236,RN257,RN3063,RN3039,RN3061,RN3062,RN3064,RN3066,Guo17}. This provides the nonequivalent lattice points in the unit cell, giving another channel for particle interaction. In many cases the second basis lowers the lattice symmetry, which changes the modes that can be excited in the system. 

One related phenomenon that appears in periodic lattices is bound states in the continuum (BICs), which are bounded (non-radiative or dark) eigenstates of an optical system despite having a dispersion within the light cone of the surrounding medium~\cite{RN3035,RN3036,RN2424,RN2771,RN2772,RN2819,RN2778,RN2774,RN2780,RN2782,RN2788,RN2789,RN2791,Abujetas2019c,Abujetas2019d,RN2779,RN2777,RN2790,RN2792,RN3168}. The quality ($Q$) factor of a BIC approaches infinity in the absence of material losses, which is of particular interest for lasing and sensing applications\cite{RN2771,Kodigala2017,Ha2018,Yesilkoy2019,Wu2020c}.  Symmetry-protected BICs often appear in high-symmetry points of the
first Brillouin zone, such as at the $\Gamma$ point. Defects and perturbations breaking the structural
symmetries allow coupling of BICs to the radiative modes,
transforming them into leaky resonances or quasi-BICs with a finite $Q$-factor.  

Since the addition of a second basis lowers the lattice symmetry, non-Bravais lattices should represent a rich field for investigating and controlling BICs. BICs in a non-Bravais lattice have been recently reported in the terahertz (THz) and visible regions using metallic systems. In the THz region, a pair of gold resonant bars (dimer) sustain anti-parallel resonances that cancel each other, forming a BIC at the $\Gamma$ point\cite{RN3037}. Such dimers have been also studied at optical frequencies, where pairs of silver nanoparticles in a lattice sustain a BIC mode\cite{RN236,RN3038,RN3039}. These works have focused on the interaction between pairs of electric dipoles, leaving unexplored the interplay with magnetic dipoles and the relation between the BICs and the symmetry of the unit cell. 

In this study, we unveil BICs emerging in non-Bravais rectangular lattices with different unit cell symmetries. The array consists of Si nanodisks which support both (in-plane and out-of-plane) electric and magnetic dipoles at telecom wavelengths. Therefore, the results presented here can be easily adaptable to any other dipolar lattice in contrast to metallic systems where only electric dipoles are excitable. The emergence of BICs is explained in an intuitive way based on the symmetry matching between the lattice and the pair of dipoles constituting the mode, which is further verified by a theoretical model based on coupled electric and magnetic dipoles (CEMD)\cite{Abujetas2018,Abujetas2020a,Abujetas2021}.

\section{non-Bravais lattice with size and position detunings}
We design the non-Bravais lattices that support BICs in the telecom region starting from a square array of Si nanodisks. The array consists of polysilicon nanodisks (200 nm height, 224 nm diameter) arranged in a square lattice with a period of $a_{\rm x}= a_{\rm y}$ = 425 nm (see Fig.\ref{fig:Figure_1}(a)). This is the initial  Bravais lattice structure that we call symmetric array. Note that the period is too short to sustain SLRs in the targeted telecom region. In the space group terminology, the symmetric array is classified as P4mm. Then we lower the symmetry of the lattice by shifting the position of the second disk by $\delta$x = 10 $\%$ from the original position (Fig.\ref{fig:Figure_1}(b)), while the diameters of the nanodisks are the same (236 nm). The size of the unit cell is doubled by this operation ($a_{x}$ = 850 nm) and the second disk lies at 2$a_{x}$/5 = 340 nm in the unit cell. It is noted that the two disks in the unit cell are not equivalent: i.e., the lattice is non-Bravais. We call this array the position-detuned array (Fig.\ref{fig:Figure_1}(b)). The space group of the position-detuned array is Pmm2, where a mirror plane lies at the center of the unit cell along the $y$-direction as indicated by a dotted line in the scanning electron microscopic (SEM) image.  Another way of lowering the symmetry is to detune the size of the disk. The diameter of the second nanodisk is thus increased by $\delta D$ = 31\% (256 nm) with respect to the first disk (194 nm). This operation also makes $a_{x}$ = 850 nm and we call it the size-detuned array (Fig.\ref{fig:Figure_1}(c)). This array also falls into Pmm2 in space group, i.e., the position- and size-detuned arrays have the same elements in terms of the symmetry.  We also make a doubly-detuned array, where both the position and the size of the second disk are detuned from the first disk by $\delta$x = 10\% and $\delta D$ = 31\%, respectively. The sample is called doubly-detuned array and the diameters of the nanodisks are 210 and 276 nm (Fig.\ref{fig:Figure_1}(d)). The space group of the doubly-detuned array is Pm; i.e., the mirror planes along the $y$-direction are lost in this array. 

\section{Results and Discussion}
\subsection{Mode dispersion at telecom frequencies}
We measure the zeroth-order (forward) transmission spectrum by illuminating the sample with a collimated beam from a supercontuniuum laser. To realize a homogeneous medium surrounding the nanodisks that provides out-of-plane symmetry, we place a quartz coverslip on top of the nanodisk array with refractive index matching oil ($\varepsilon=2.1$) in between. We evaluate the optical extinction ($E$), defined as $E=1-T/T_{0}$, where $T$ is the transmission through the sample and $T_{0}$ is the transmission through a reference consisting of a substrate, index-matching oil and superstrate. The extinction spectra are shown for TE-polarization in Figs.\ref{fig:Figure_1}(e)-(h) as a function of the angle of incidence, $\theta_{\rm in}$, where the incident plane is defined in the $z-x$ plane. 

For the symmetric array (Fig.\ref{fig:Figure_1}(e)), no extinction features appear in the range of measurement ($\lambda$ between 1100 and 1500 nm) because the lattice resonances occur in the visible frequency range. On the other hand, for the position-detuned array (Fig.\ref{fig:Figure_1}(f)), highly dispersive features appear following the in-plane diffraction orders (Rayleigh anomalies), which correspond to the SLRs. The Rayleigh anomalies satisfy 
\begin{equation}
\vec{k}_{0}=\vec{k}_{x}(\theta_{\rm {in}}) \pm \vec{G}(m_{1},m_{2}),
\end{equation}
where $\vec{k}_{0}$ is the wave vector of the in-plane diffracted orders, $\vec{k_x}$ is the in-plane component of the wave vector of the incident beam, and $\vec{G}(m_{1},m_{2}) = m_{1}(\frac{2\pi}{a_{x}})\vec{x}+m_{2}(\frac{2\pi}{a_{y}})\vec{y}$ 
is the reciprocal lattice vector of the array with $m_{1}$ and $m_{2}$ the orders of diffraction in the $x-$ and $y-$ directions, respectively. 
At the $\Gamma$ point, the SLRs bend with a splitting energy of $\sim$ 41 meV between the upper and lower modes.  The longer wavelength or lower-energy SLR becomes narrower as $\theta_{\rm in}$ approaches 0$^\circ$, leading to a BIC at $\theta_{\rm in}$ = 0$^\circ$ as seen in Fig.\ref{fig:Figure_1}(f). 
SLRs also appear for the size-detuned array (Fig.\ref{fig:Figure_1}(g)), and interestingly enough, the BIC mode corresponds to the higher-energy SLR. 
For the doubly-detuned array (Fig. \ref{fig:Figure_1}(h)), both SLRs are bright. The cuts to the extinction dispersion measurements further show the evolution of the SLRs around the $\Gamma$ point (Figs. \ref{fig:Figure_1}(i-l)). 

The emergence of these BICs at the $\Gamma$ point can be understood in an intuitive way upon considering the modes supported by the symmetric array and how they evolve as the symmetry of the lattice is changed, as depicted in Figs. \ref{fig:Figure_2}(a-b). In the frequency range of interest, the symmetric array supports two guided modes given by the out-of-phase (anti-parallel) oscillation of in-plane electric dipoles (ED$_{\rm y}$) and out-of-plane magnetic dipoles (MD$_{\rm z}$). We will refer to them as the electric and magnetic guided modes, respectively. Since guided modes cannot be excited with propagating plane waves, the extinction spectrum shows no features (see Figs. \ref{fig:Figure_1}(e) and (i)). In contrast, for the other arrays, the perturbation of the unit cell doubles their size along the $x$ axis and the dispersion relations of the guided modes fold down to the continuum of radiation. Therefore, the initially guided modes of the symmetric array evolve to leaky (bright) resonances or even to BICs. 

For the position-detuned array,
the displacement of the second disk in each (new) unit cell allows the coupling between ED$_{\rm y}$ and MD$_{\rm z}$, forming hybrid modes. At the $\Gamma$ point, the eigenstate stemming from the magnetic (electric) guided mode is given by anti-parallel oscillations of MD$_{\rm z}$ (ED$_{\rm y}$) mixed with parallel oscillations of ED$_{\rm y}$ (MD$_{\rm z}$), as depicted in Fig. \ref{fig:Figure_2}(a) (2(b)). Note that both, ED$_{\rm y}$ and MD$_{\rm z}$, oscillating parallel or anti-parallel are not self-consistent dipole configurations in the position-detuned array. Since the MD$_{\rm z}$ does not radiate in any case at the $\Gamma$ point\cite{RN3074} and the anti-parallel ED$_{\rm y}$ cancels with each other, the electric guided mode evolves to a BIC (Fig. \ref{fig:Figure_2}(b)). In contrast, the magnetic guided mode evolves to a bright mode, due to the parallel oscillations of the ED$_{\rm y}$ associated to this eigenmode (Fig. \ref{fig:Figure_2}(a)). On the other hand, for the size-detuned array, ED$_{\rm y}$ and MD$_{\rm z}$ do not mix, but the relative strength between the dipoles in each cell is different. Therefore, the magnetic guided mode evolves to a BIC (MD$_{\rm z}$ does not radiate at the $\Gamma$ point), while the electric guided mode evolves to a bright resonance, since the unequal amplitudes of the ED$_{\rm y}$ break the symmetry at normal incidence as seen in Figs. \ref{fig:Figure_2}(a) and (b), respectively. 
Finally, for the doubly-detuned array, both resonances can couple to a plane wave at the $\Gamma$ point, so that both become bright. This is understood by combining the phenomenology exhibited for the position- and size-detuned arrays.  The absence of symmetry-protected BICs is in accordance with the fact that the doubly-detuned array shows the lowest symmetry among the arrays in the present study. 

The cuts to extinction spectra at normal incidence are shown in Figs. \ref{fig:Figure_2}(c-f), together with the schematic of the modes supported by each lattice. The phenomenology described in terms of the eigenmodes of the systems adequately suits with the experimental results: The modes related to the magnetic guided mode are always at higher energies than those derived from the electric guided mode. Also, the mode frequency is only slightly altered upon a lattice perturbation (note that BIC frequencies can be estimated from the extinction dispersion spectra), suggesting that the framework depicted in Figs. \ref{fig:Figure_2}(a) and (b) is correct.

For the sake of completeness, the extinction for TM light is shown in Supporting Fig. S1. The two SLRs also appear for TM polarization, but the splitting is smaller ($\sim$ 21 meV). For this polarization, in-plane MDs instead of EDs couple to the diffraction order\cite{RN0001,RN2889,RN3073}, and a smaller splitting indicates a smaller in-plane magnetic polarizability of the Si nanodisk compared to the electric polarizability. For the position-detuned array, the BIC from anti-parallel MD$_{\rm y}$ mixed with parallel ED$_{\rm z}$ is supported, while anti-parallel ED$_{\rm z}$ creates the BIC for the size-detuned array. 

\subsection{CEMD calculation}
To confirm our intuitive explanation of the underlying physical mechanism, we have investigated theoretically the extinction of the arrays using a CEMD model.~\cite{Abujetas2018a,Abujetas2020a}
We extract numerically the polarizability of electric and magnetic dipoles from the single disks, and calculate analytically $1-T$ (equivalent to the experimental extinction, since $T$ is the specular transmittance) when the dipoles are arranged in an infinite lattice.  The calculations shown in Figs. \ref{fig:Figure_3}(a-d) reproduce fairly well the experimental results in Figs. \ref{fig:Figure_1}(e-h). No modes appear for the symmetric array as in the experiment. The flip of energy of the bright and BIC modes in position- and size-detuned arrays is clearly seen in Figs. \ref{fig:Figure_3}(b) and (c), and the two bright modes at the $\Gamma$ point are obvious for the doubly-detuned array (Fig. \ref{fig:Figure_3}(d)). 

We further verify the intuitive assignment of the modes in Fig.\ref{fig:Figure_2} by projecting the total calculated extinction into the in-plane and out-of-plane ED and MD components.  The bottom panels in Fig. \ref{fig:Figure_3} show the cuts to the extinction spectra around $\theta_{\rm in}$ = 0$^\circ$. The color of the lines for all three detuned arrays in Figs.~\ref{fig:Figure_3}(b)-(d) reflects the origin of the modes, in accordance with the assignments in Fig.~\ref{fig:Figure_2} for the lower and higher energy bands:  red represents the projection over the low-energy state with anti-parallel ED$_{y}$ mixed with parallel MD$_{z}$, while blue denotes the projection over the high-energy state with anti-parallel MD$_{z}$ mixed with parallel ED$_{y}$. This projection is done because these states are the theoretical eigenstates of the system at normal incidence.
The brightness of the mode is determined by the position and magnitude of the dipoles, which are governed by the position and size of the disks, respectively, as explained in Fig. \ref{fig:Figure_2}. The  projections of the theoretical results support the above-mentioned nature of the bands and the emergence of BICs at the $\Gamma$ point in Figs. \ref{fig:Figure_3}(f) and (g), only plausible with combinations of anti-parallel (identical) in-plane ED and/or arbitrary vertical MDs. Away from the $\Gamma$ point the SLRs bands  may exhibit more complex combinations of dipolar states, as revealed in Figs. \ref{fig:Figure_3}(f-h) by the mixture of red/blue colors.
We also plot separately the contributions from the ED$_{y}$ and MD$_{z}$ components to the total extinction as a function of frequency and angle of incidence in Supporting Fig. S2; note that the combination of both in-plane ED and out-of-plane MD contributions are needed to reproduce the discussed SLR bands and resulting BICs.
 
\begin{figure}
\includegraphics[width=0.9\textwidth]{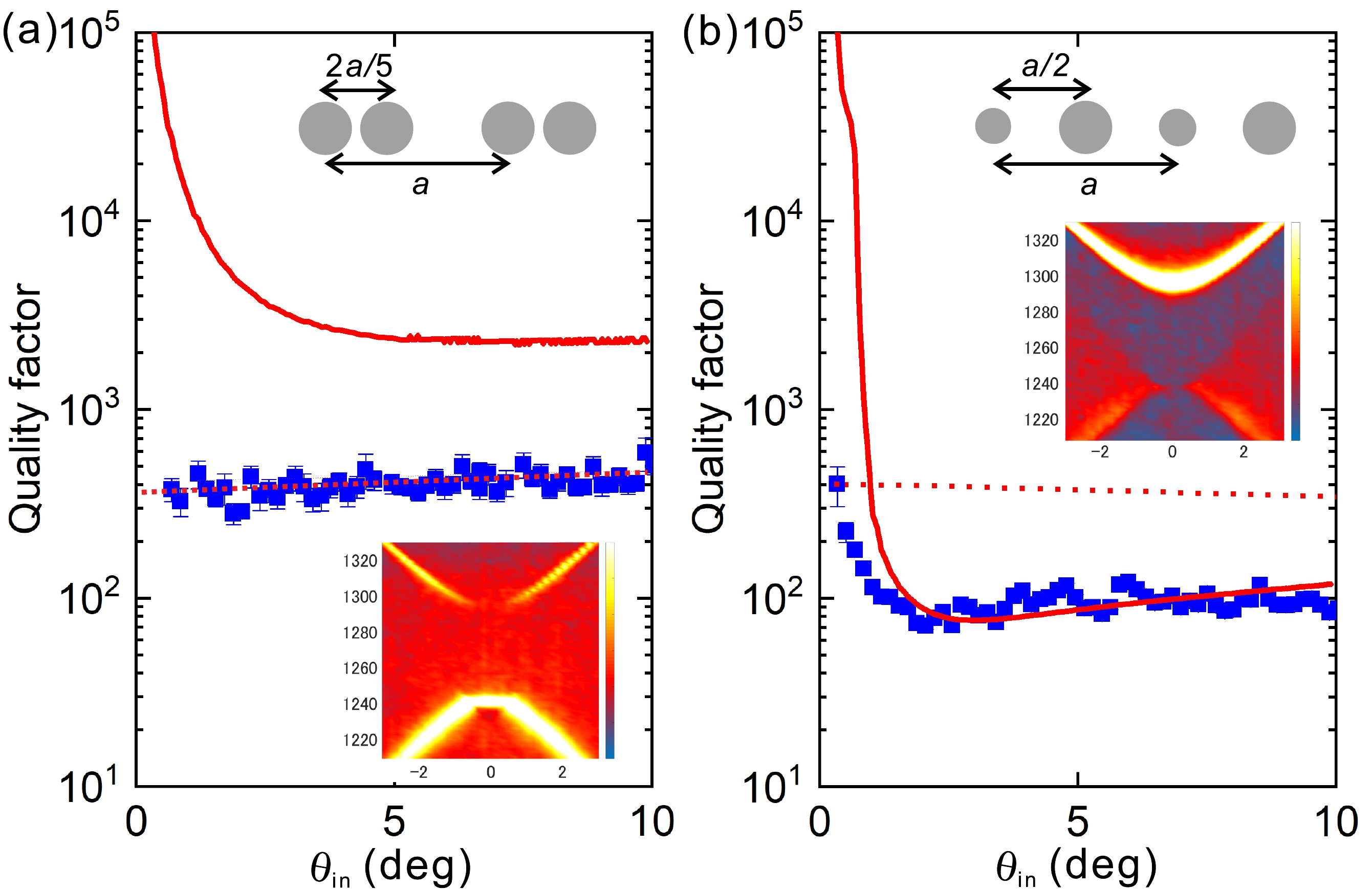}
\caption{$Q$-factors of quasi-BIC mode extracted from the extinction as a function of $\theta_{\rm in}$ for TE polarized light. (a) the position-detuned array, and (b) the size-detuned array. The dotted line indicate the upper limit of the $Q$-factor imposed by the resolution of the spectrometer. The $Q$-factors extracted from the CEMD calculation in Figure 3 are shown as solid lines. The upper and lower insets are the schematic representation of the array and the optical extinction around normal incidence.}
\label{fig:Figure_4}
\end{figure}

The direct evidence of the transition from quasi-BIC to symmetry-protected BIC is the resonance narrowing as it approaches $\theta_{\rm in}$ = 0 $^\circ$. We extract the $Q$-factors from the two quasi-BICs appearing for the position- and size-detuned arrays in Fig. \ref{fig:Figure_4}. For the position-detuned array in Fig.~\ref{fig:Figure_4}(a), the $Q$-factor is so large that the value is above the resolution limit of the spectrometer (indicated by the dotted line in the plot) for all the measured angles. We also have plotted the $Q$-values extracted from the CEMD calculation by solid lines, showing a diverging behaviour as $\theta_{\rm in}$ approaches 0$^\circ$. For the size-detuned array (Fig.~\ref{fig:Figure_4}(b)), the $Q$-factor increases as $\theta_{\rm in}$ approaches 0 $^\circ$ and reaches the detection limit imposed by the resolution of the spectrometer. The agreement between the experiment and calculation is excellent. This result might be surprising given the rough surface of the nanoparticles as observed in the SEM images shown in Fig.~\ref{fig:Figure_1}. One possible reason for the excellent agreement between experiments and calculations is the working 
wavelength in the present work: Telecom wavelengths are relatively long compared to the surface imperfections and scattering can be neglected. Also the absorption of Si is negligible in this wavelength range, allowing the disks to be regarded as pure magnetic and electric dipoles.

\subsection{Spectral tuning of the BIC position}
In order to demonstrate the universality of the BICs appearing in rectangular non-Bravais lattice, we change the parameters in the design to shift the features. When the size of nanodisks is enlarged while the lattice constants remain the same, the two bands redshift with a larger splitting ($\sim$ 83 meV for TE light) due to the larger polarizability of the disks.  Details are found in Supporting Sec. 3.

\begin{figure}
\includegraphics[width=1\textwidth]{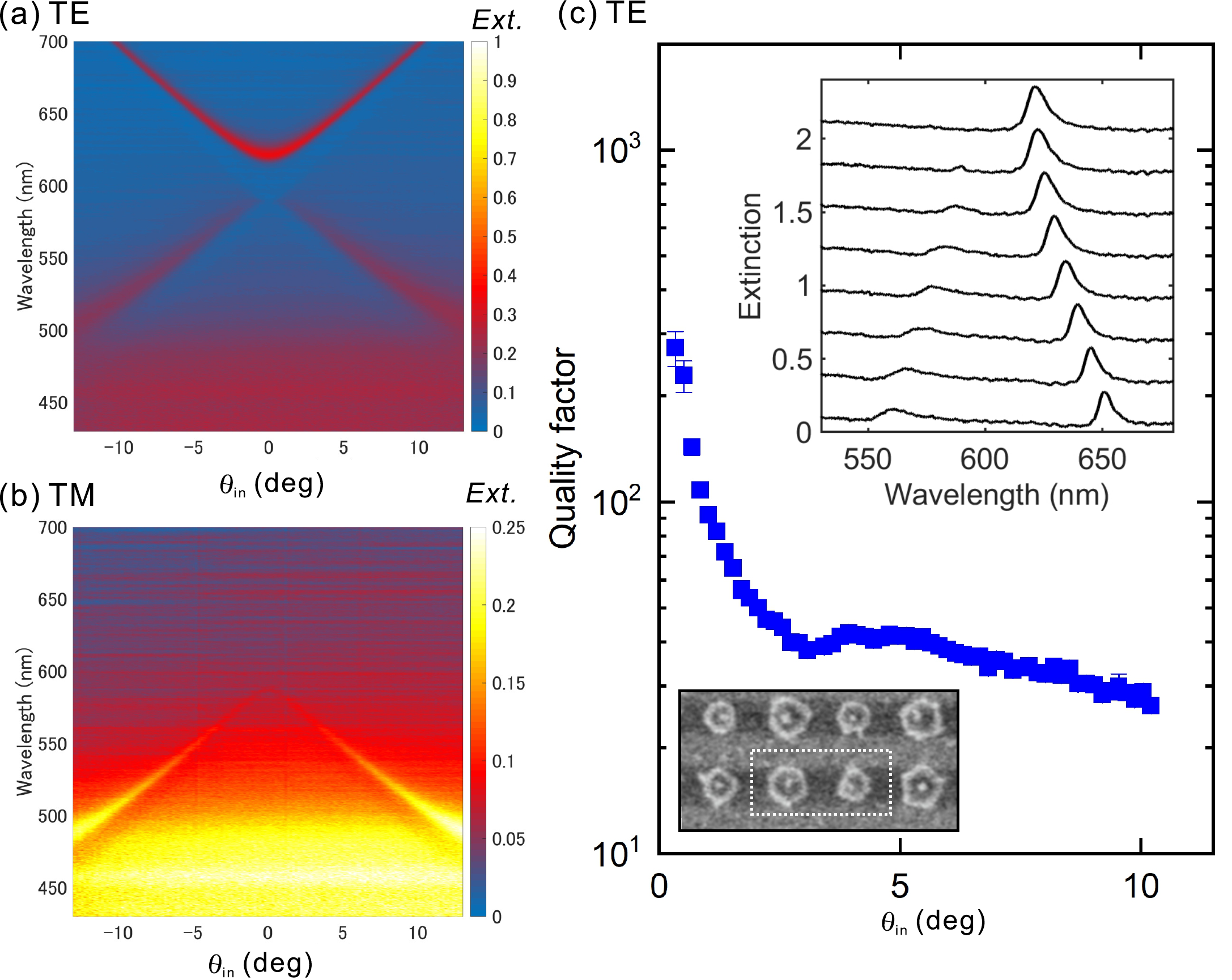}
\caption{Optical properties of the size-detuned array resonating in the visible. The array consists of the pair of Si nanodisks (200 nm height, 86 and 107 nm diameters, center-to-center distance 200 nm) are arranged in the $x$- and $y$-directions with the period of 400 and 200 nm, respectively. (a) and (b): Extinction spectra as a function of $\theta_{\rm in}$ for (a) TE and (b) TM polarized light. (c) $Q$-factors of quasi-BIC mode extracted from the extinction as a function of $\theta_{\rm in}$ for TE-polarized light. The top inset is the extinction spectra around normal incidence ($\theta_{\rm in}$ between 0 to 5.14$^\circ$ every 0.68 $^\circ$ step). The spectra are shifted vertically for the sake of clarity.
The bottom inset is a top-view SEM image of the array. The dotted rectangle denotes the unit cell (400 nm $\times$ 200 nm) of the array.}
\label{fig:Figure_5}
\end{figure}

We further shift the spectral features to the visible range by scaling the structure. The prepared size-detuned array consists of polysilicon nanodisks (86 and 107 nm diameters) arranged with $a_x$ = 400 nm and $a_y$ = 200 nm (see the inset in Fig.~\ref{fig:Figure_5}(c) for the SEM image). The extinction shows a localized dispersion-less band at $\lambda$ $\leq$ 500 nm and the two SLRs (Fig.~\ref{fig:Figure_5}). At $\theta_{\rm in}$=0$^\circ$, the high and low-energy bands appear with an energy gap of $\sim$ 105 meV for TE light (Fig.~\ref{fig:Figure_5}(a)), and the higher-energy SLR becomes a BIC. For TM polarization (Fig.~\ref{fig:Figure_5}(b)), the BIC mode is flipped to the lower energy band although less obvious due to a smaller magnetic polarizability. The appearance of BIC is evidenced by the $Q$-factor as a function of  $\theta_{\rm in}$, where $Q$ diverges as $\theta_{\rm in}$ approaches 0$^\circ$ in Fig.~\ref{fig:Figure_5}(c).

\section{Conclusions}
We have demonstrated experimentally the emergence of symmetry-protected BICs at telecom wavelengths by detuning square arrays of polysilicon nanodisks into rectangular non-Bravais lattices. The initial square array supports two guided modes, comprising anti-parallel EDs and MDs, respectively, that turn into SLRs upon doubling the unit cell length in one direction by modifying the position/size of every other disk. For the position-detuned array, the TE (TM) mode polarized along the short (long) axis of the lattice, consisting of horizontal anti-parallel EDs (MDs) mixes with vertical parallel MDs (EDs), becoming a BIC; for the size-detuned array, conversely, vertical anti-parallel MDs (EDs) lead to a BIC for the TE (TM) mode. 
When the symmetry of the lattice is further lowered by detuning size and position of the second nanodisks simultaneously, the symmetry-protected BICs disappear.
The measurements of the $Q$-factor of the BIC is limited by the resolution of the spectrometer, being this $Q$-factor larger than 400. This work provides a robust design strategy of symmetry-protected BICs via introduction of a second meta-atom (nanodisk in this case) per unit cell to Bravais lattices, which is readily applicable to lasing, sensing, and light-matter coupling researches. 
\section{Methods}
\textbf{Fabrication.} Polycrystalline Si thin films with a thickness of 80 nm were grown on a synthetic silica glass substrate by low-pressure chemical vapor deposition using SiH$_4$ gas as a source of Si. A resist (NEB22A2, Sumitomo) was cast on the Si film and exposed to electron-beam lithography, followed by development to make nanoparticle arrays of resist on the Si film. The Si film was vertically etched using a selective dry etching (Bosch process) with SF$_6$ and C$_4$H$_8$ gases, and the resist residue was etched away by oxygen dry etching. The fabricated area of the array covers 3 $\times$ 3 mm$^2$.

\begin{figure}
\includegraphics[width=1\textwidth]{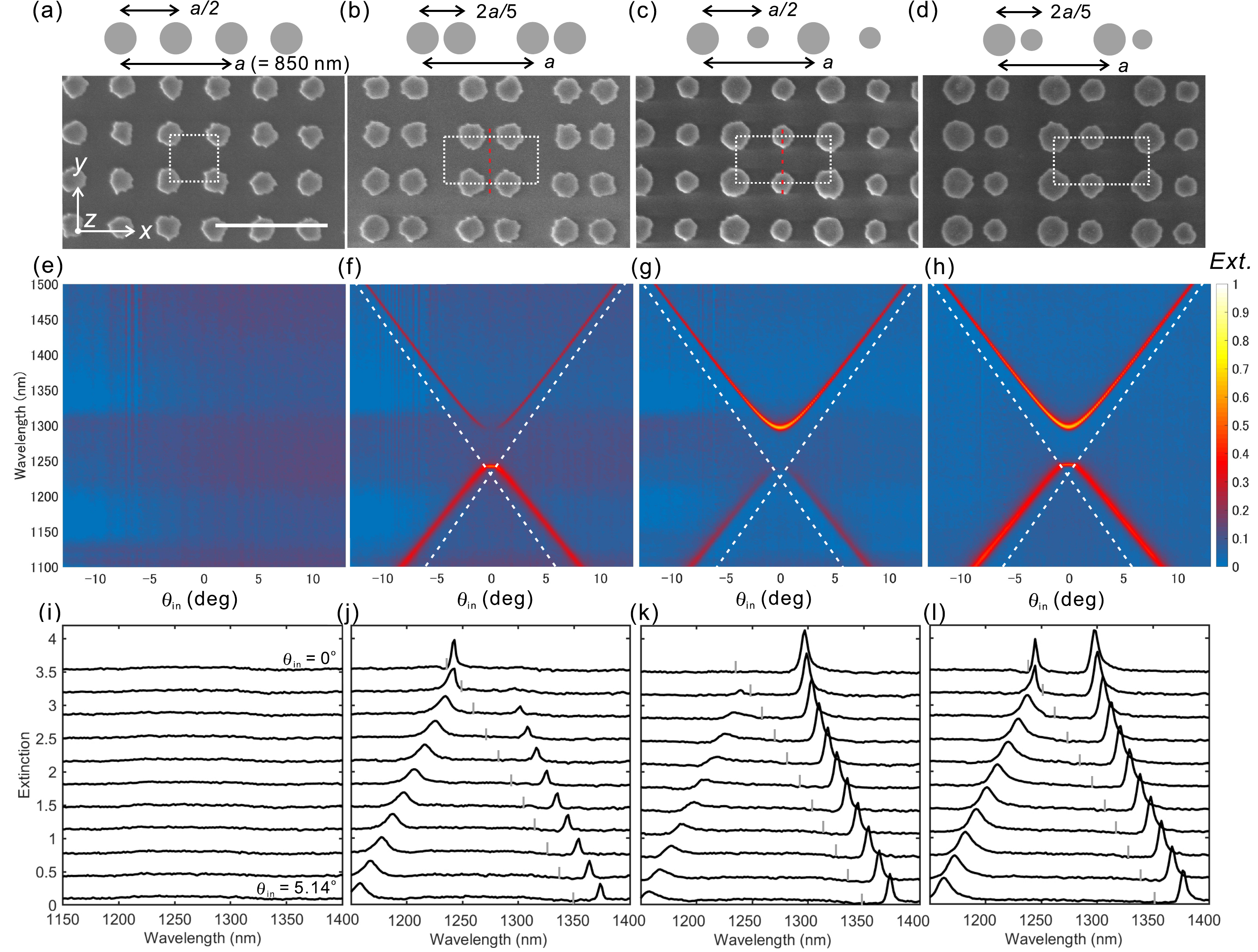}
\caption{Structure and extinction of the nanoparticle arrays. SEM images of (a) the symmetric array, where Si nanodisks (200 nm height, 224 $\pm$ 7.2 nm diameter) are arranged in the $x$- and $y$-directions with the period of 425 nm, (b) the position-detuned array, where the pair of Si nanodisks (200 nm height, 236 $\pm$ 6.6 nm diameter, center-to-center distance 340 nm) are arranged in the $x$- and $y$-directions with the period of 850 and 425 nm, respectively, and (c) the size-detuned array, where the pair of Si nanodisks (200 nm height, 194 $\pm$ 3.6 and 256 $\pm$ 6.6 nm diameters, center-to-center distance 425 nm) are arranged in the $x$- and $y$-directions with the period of 850 and 425 nm, respectively, and (d) doubly-detuned array, where the pair of Si nanodisks (200 nm height, 210 $\pm$ 7.0 and 276 $\pm$ 2.6 nm diameters, center-to-center distance 340 nm) are arranged in the $x$- and $y$-directions with the period of 850 and 425 nm, respectively. The unit cell is denoted as white dashed boxes. The sketches of the disk arrangement along the $x$-direction are on the top. Extinction spectra (1-zeroth-order transmittance) for TE light as a function of angle of incidence $\theta_{\rm in}$ for the (e) symmetric array, (f) position-detuned array, (g) size-detuned array, and (h) doubly-detuned array. The in-plane diffraction condition is denoted as dotted lines. Extinction spectra (TE) around the normal incidence ($\theta_{\rm in}$ between 0 and 5.14 $^\circ$ in steps of 0.514 $^\circ$ step) and the mode assignment: (i) the symmetric array, (j) the size-detuned array, (k) the position-detuned array, and (l) the doubly-detuned array. The in-plane diffraction condition is denoted by vertical ticks. } 
\label{fig:Figure_1}
\end{figure}

\begin{figure}
\includegraphics[width=1\textwidth]{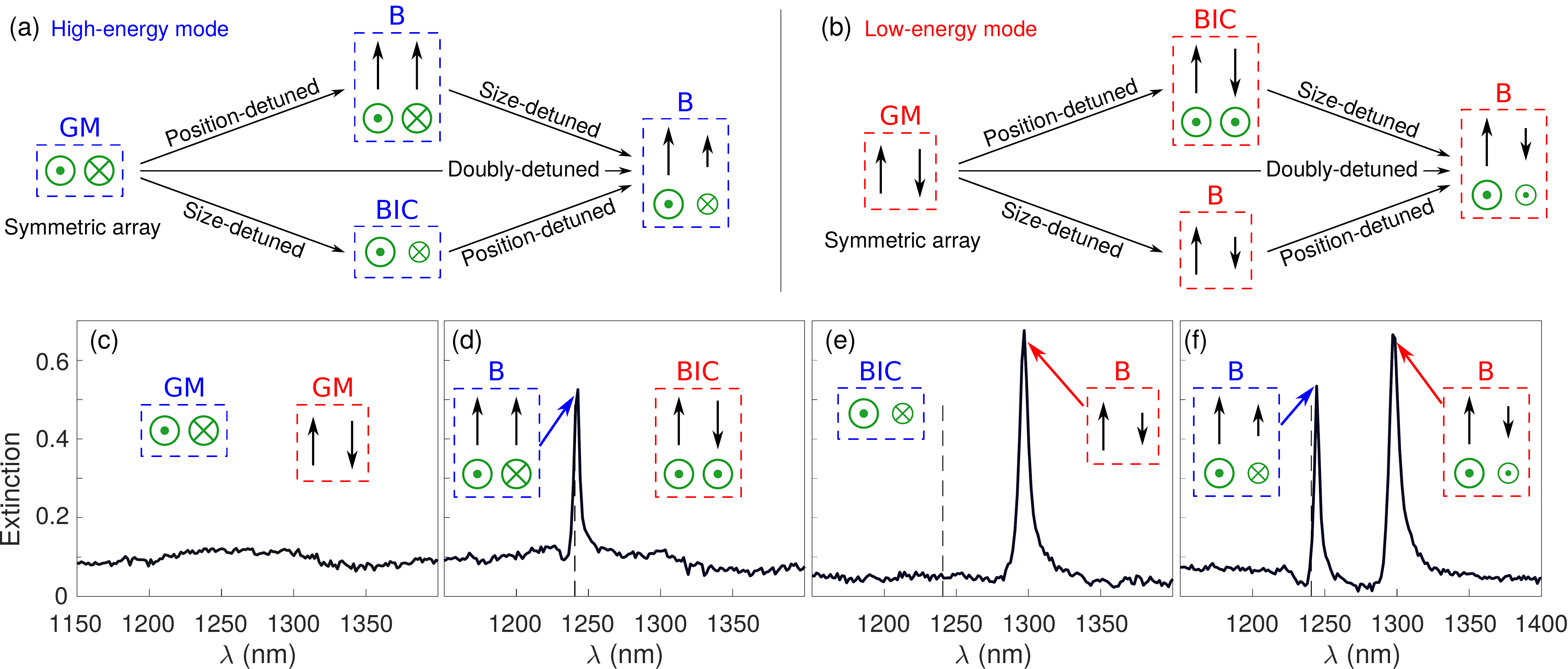}
\caption{Schematic of mode evolution under lattice modification and correlation with extinction spectra (TE) at normal incidence ($\theta_{\rm in}$ = 0$^\circ$). (a) Evolution of the out-of-plane magnetic dipoles and (b) of the in-plane electric dipoles eigenmodes from the guided modes in symmetric arrays to bright and BIC modes in other lattice symmetries. GM: guided mode, B: bright mode. (c-f) Extinction spectra (TE) at normal incidence taken from Fig. \ref{fig:Figure_1} (i-l). (c) Symmetric, (d) position-detuned, (e) size-detuned and (f) double-detuned arrays. The insets show the eigenmodes supported by each lattice symmetry and the vertical dashed lines indicate the in-plane diffraction condition.}
\label{fig:Figure_2}
\end{figure}

\begin{figure}
\includegraphics[width=1\textwidth]{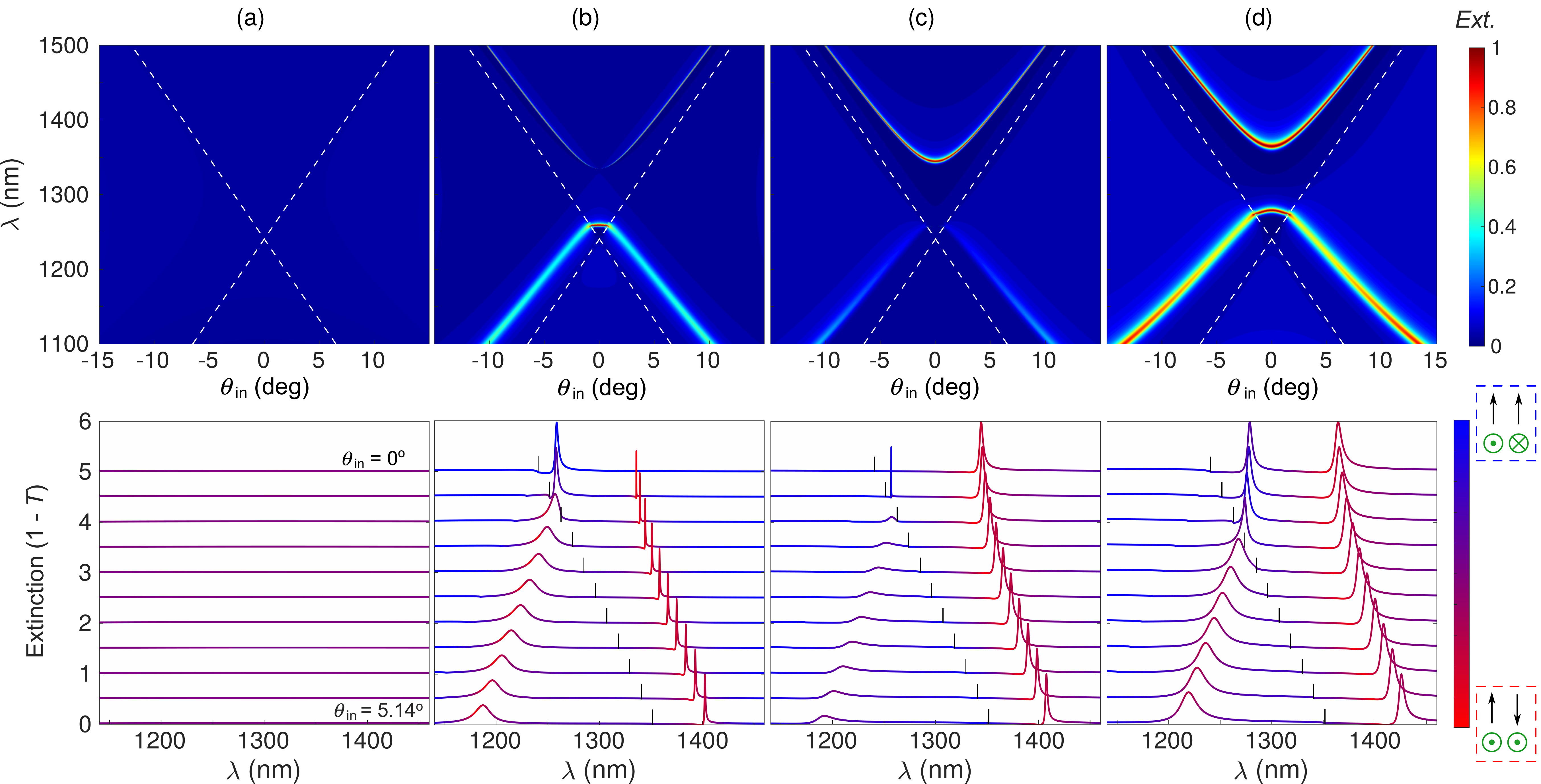}
\caption{Extinction spectra ($1-T$, $T$ being the specular transmittance) of the nanoparticle arrays calculated using the CEMD method for: (a) a symmetric array, where Si nanodisks (200 nm height, 225 nm diameter) are arranged in a square lattice with a period of 425 nm; (b) a position-detuned array, where Si nanodisk dimers (200 nm height, 225 nm diameter, center-to-center distance 340 nm) are arranged in a rectangular lattice with periods of 850 and 425 nm; (c) a size-detuned array, where Si nanodisk dimers (200 nm height, 200 and 250 nm diameters, center-to-center distance 425 nm) are arranged in a rectangular lattice with periods of 850 and 425 nm;  and (d) a doubly-detuned array, where Si nanodisk dimers (200 nm height, 210 and 270 nm diameters, center-to-center distance 340 nm) are arranged in a rectangular lattice with periods of 850 and 425 nm, respectively. Upper panels show the extinction spectra (TE) as a function of $\theta_{\rm in}$. (e-h) The bottom panels are the spectra shown in (a-d) between $\theta_{\rm in}$= 0$^\circ$ and 5.14$^\circ$ in steps of 0.514$^\circ$. The color of the curves denotes the projection over the collective modes shown in the insets in Figs.~\protect{\ref{fig:Figure_2}}(b-d):  (red)  low-energy mode with anti-parallel pairs of ED$_{y}$ and parallel MD$_{z}$; and (blue) high-energy mode with parallel ED$_{y}$ and anti-parallel MD$_{z}$. The non-diffraction region is delimited by the vertical ticks.}
\label{fig:Figure_3}
\end{figure}

\textbf{Extinction measurements.}
The nanodisk arrays were in an optically homogeneous environment by placing an upperstrate of synthetic silica glass with index-matching oil. The samples were placed in a rotational stage, and illuminated by a collimated and polarized beam. The source of illumination was a broadband supercontinuum laser (FIANIUM WHITELASE MICRO). The beam diameter was $0.1$ mm, smaller than the area of the array. The optical axis corresponds to the $z-$axis and the sample was rotated around the $y-$axis (see Fig. \ref{fig:Figure_1}a). The polarization of the incident light was fixed either along the $y$-axis (TE polarization) or the $x$-axis (TM polarization).  As the sample was rotated along the $y$-axis by an angle $\theta_{\rm in}$, light was incident with a wave vector component that was parallel to the array plane along the $x$-direction, $k_x=\frac{2\pi n}{\lambda}\sin{\theta_{\rm in}}$, where $\lambda$ is the free-space wavelength and $n$ is refractive index of the glass. We measured the spectra with a fiber-coupled spectrometer (NIR Quest, Ocean Optics) with 3.1 nm spectral resolution. We calculated the extinction as $E = 1 - T/T_{0}$, where $T$ and $T_{0}$ are the zeroth-order transmission measured on and off the array, respectively. 

\textbf{CEMD theoretical calculations.}
We make use of our coupled electric/magnetic dipole theory (CEMD) for infinite 2D planar arrays developed in Refs.~\cite{Abujetas2018,Abujetas2020a,Abujetas2021}. In the case of disk meta-atoms with axes perpendicular to the metasurface, as explained in Ref.~\cite{Murai2020c}, we need the polarizabilities of the in-plane and out-of-plane, electric and magnetic resonances to apply the CEMD theory. These polarizabilities are numerically obtained through SCUFF \cite{SCUFF1,SCUFF2}, an open-source software package for analysis of electromagnetic scattering problems using the method of moments, the dielectric function of polycrystalline silicon $\varepsilon=3.5$,\cite{RN2425,RN3071,RN3072} and assuming a medium surrounding the nanodisks with $\varepsilon=2.1$. The  four calculated polarizibilites are introduced in our CEMD model, i.e., in- and out-of-plane MD and ED, which accounts for all lattice-induced interactions.

\begin{acknowledgement}

This work was partly supported by the Nanotechnology Hub, Kyoto University and Kitakyusyu FAIS in the “Nanotechnology Platform Project,” sponsored by MEXT, Japan. We gratefully acknowledge the financial support from Kakenhi (17KK0133, 19H02434) and the Netherlands Organisation for Scientific Research (NWO) through Gravitation grant “Research Centre for Integrated Nanophotonics” and Innovational Research Activities Scheme (Vici project SCOPE no. 680-47-628). J.A.S.-G. and D.R.A. acknowledge support from the Spanish Ministerio de Ciencia e Innovaci\'on through grants MELODIA PGC2018-095777-B-C21 and FPU PhD Fellowship FPU15/03566 (MCIU/AEI/FEDER, UE). 

\end{acknowledgement}
\begin{suppinfo}
Supporting Information is available:
Extinction for the TM-polarized light; Contributions of in- and out-of-plane components to the extinction; BICs in the size-detuned array with larger disk diameters.

\end{suppinfo}
\bibliography{library}
\newpage

\newpage


\newpage

\newpage

\newpage
\makeatletter

\end{document}